\documentclass[english,aps,superscriptaddress,twocolumn]{revtex4}
\usepackage[T1]{fontenc}
\usepackage[utf8]{inputenc}
\usepackage{amsbsy}
\usepackage{graphicx}

\usepackage{color}

\usepackage{babel}

\begin{document}

\title{Baryon formation and dissociation in dense hadronic and quark matter}

\author{Jin-cheng Wang}

\affiliation{Interdisciplinary Center for Theoretical Study and Department of
Modern Physics, University of Science and Technology of China, Anhui
230026, People's Republic of China}

\affiliation{Institute for Theoretical Physics, Johann Wolfgang Goethe
  University, Max-von-Laue-Str.\ 1, D-60438 Frankfurt am Main, Germany}

\author{Qun Wang}

\affiliation{Interdisciplinary Center for Theoretical Study and Department of
Modern Physics, University of Science and Technology of China, Anhui
230026, People's Republic of China}

\affiliation{Theoretical Physics Center for Science Facilities, Chinese Academy
of Sciences, Beijing 100049, People's Republic of China}

\author{Dirk H.\ Rischke}

\affiliation{Institute for Theoretical Physics, Johann Wolfgang
Goethe
  University, Max-von-Laue-Str.\ 1, D-60438 Frankfurt am Main, Germany}

\affiliation{Frankfurt Institute for Advanced Studies,
Ruth-Moufang-Str.\ 1, D-60438 Frankfurt am Main, Germany}

\begin{abstract}
We study the formation of baryons as composed of quarks and diquarks
in hot and dense hadronic matter in a Nambu--Jona-Lasinio (NJL)--type model.
We first solve the Dyson-Schwinger equation for the diquark propagator
and then use this to solve the Dyson-Schwinger equation for the
baryon propagator. We find that stable baryon resonances exist only in
the phase of broken chiral symmetry. In the chirally symmetric phase,
we do not find a pole in the baryon propagator.
In the color-superconducting phase, there is a pole, but is has
a large decay width. The diquark does not need to be stable in order to
form a stable baryon, a feature typical for so-called Borromean states.
Varying the strength of the diquark coupling constant,
we also find similarities to the properties of an Efimov states.
\end{abstract}

\maketitle

A baryon is a color-singlet bound state of three constituent quarks.
Since the interaction between two quarks is attractive in the
color-antitriplet channel, baryon formation can be regarded as
a two-step process: first, two quarks combine to form
a diquark with color-antitriplet quantum numbers, and
then this diquark combines with another color-triplet quark to form a
color-singlet bound state \cite{GellMann:1964nj,Ida:1966ev,
Lichtenberg:1967zz,Efimov:1990uz,Anselmino:1992vg,Buck:1992wz,Ishii:1993rt,AbuRaddad:2002pw,Zou:2005xy}.

At extremely high baryonic densities and low temperatures quarks
form Cooper pairs in the attractive color-antitriplet channel,
leading to the phenomenon of color superconductivity
\cite{Alford:1997zt,Rapp:1997zu,Pisarski:1999bf,Hong:1999fh} [for
recent reviews, see e.g.\ Refs.\ \cite{Alford:2007xm,Wang:2009xf}]. Because
of asymptotic freedom, the interaction is weak and, just like in BCS
theory, the Cooper pair wave function has a correlation length that
exceeds the interparticle distance. However, as the density is
lowered, the interaction strength increases and the Cooper pair
becomes more and more localized \cite{Abuki:2001be,Abuki:2006dv}.
Eventually, Cooper pairs will form tightly bound molecular diquark
states \cite{Kitazawa:2007zs}. These may pick up another quark with
the right color to form a color-singlet baryon. This is what must
happen across the deconfinement transition into the hadronic phase.
Understanding the nature of the transition between dense hadronic
and quark matter is one of the scientific goals of the Compressed
Baryonic Matter (CBM) experiment planned at the Facility for
Antiproton and Ion Research (FAIR) \cite{Senger:2009zz}.

In this paper we investigate the formation and dissociation of
baryons in different regions of
the phase diagram of strongly interacting matter: the phase of broken
chiral symmetry (hadronic phase), the phase of restored chiral symmetry
(the quark-gluon plasma) above and below the dissociation boundary for
diquarks, and the phase where quark matter is a color superconductor.
We use an NJL-type model
\cite{Hatsuda:1994pi,Buballa:2003qv} for two quark flavors
and employ the following strategy. First, we compute the
full propagator for the scalar diquark state
via solving a Dyson-Schwinger equation.
With the diquark propagator and an additional quark propagator, we then
solve a Dyson-Schwinger equation for the baryon propagator.

Our approach bears some similarities to previous studies of diquark and
baryon formation
\cite{Ishii:1995bu,Pepin:1999hs,Bentz:2001vc,Bentz:2002um,Gastineau:2005wm}.
These works also considered an NJL-type model, but they solved the full
Faddeev equation instead of a (simpler) Dyson-Schwinger equation to obtain
baryon states. The difference is that in the Faddeev equation the coupling
between quark and diquark is not assumed to be local: a non-static
quark can be exchanged between them. Our work
is based on the cruder approximation of a local quark-diquark coupling.
These works also considered the
axial-vector diquark state, not only the scalar one,
and thus were able to investigate also excited baryon states.
On the other hand, in those works only the zero-temperature case was studied, while we also
consider non-zero temperature. Moreover, we do not assume the
diquark to be a well-defined quasi-particle
in order to solve the Dyson-Schwinger equation (an approximation employed
in the aforementioned works in order to solve the Fadeev equation).
We shall see that diquarks can also be unstable, but still give
rise to stable baryons, a typical feature of a Borromean state
also encountered in atomic and nuclear physics.
Varying the diquark coupling strength, we also find that our baryon
has properties which bear similarities to those of an Efimov state.
We use natural units $\hbar=c=k_B=1$; the metric tensor is
$g_{\mu \nu} = {\rm diag} (+,-,-,-)$.

The Lagrangian of the two-flavor NJL model with diquark-diquark
interactions reads
\begin{eqnarray}
\mathcal{L}_{NJL} & = &
\overline{\psi}(i\gamma_{\mu}\partial^{\mu}-\hat{m}_{0}
+\hat{\mu}\gamma_0 )\psi \nonumber \\
&& +G_{S}[(\overline{\psi}\psi)^{2}+
(\overline{\psi}i\gamma_{5}\boldsymbol{\tau}\psi)^{2}]\nonumber \\
 &  & +G_{D}[\overline{\psi}i\gamma_{5}\tau_{2}J_{a}\psi_{C}]
[\overline{\psi}_{C}i\gamma_{5}\tau_{2}J_{a}\psi]\;.
\label{eq:njl01}\end{eqnarray}
Here, we have suppressed the color indices in the fundamental
representation, $a=1,2,3$, and the flavor indices, $\alpha=u,d$,
in the quark spinors $\psi\equiv\psi_{a\alpha}$. The bare mass matrix is
$\hat{m}_{0}=\mathrm{diag}(m_{u}^{(0)},m_{d}^{(0)})$
and the chemical potential matrix is $\hat{\mu}=\mathrm{diag}(\mu_{u},\mu_{d})$,
$\tau_{s}$ ($s=1,2,3$) are the Pauli matrices in flavor space,
$(J_{a})_{bc}=-i\epsilon_{abc}$
are the antisymmetric color matrices, $G_{S}$ and $G_{D}$
are coupling constants for quark-antiquark and quark-quark interactions,
respectively. In principle,
$G_D$ can be related to $G_S$ via a Fierz transformation, but
we choose to keep it as a free parameter, allowing to explore
a wider range of potentially interesting phenomena within our
effective model for the strong interaction.

In the following, we neglect
the contribution from the isovector quark-antiquark channel,
$\overline{\psi}i\gamma_{5}\boldsymbol{\tau}\psi = 0$. 
We also decompose the scalar quark current in terms of a condensate part
and a fluctuation,
$\bar{\psi}_\alpha \psi_\alpha = \sigma_\alpha +
\delta_\alpha$, where $\sigma_{\alpha}= \left\langle
\overline{\psi}_{\alpha}\psi_{\alpha}\right\rangle $ is
the chiral condensate, and we work in the mean-field approximation,
i.e., we neglect terms of order $O(\delta_\alpha^2)$.
Similarly, we decompose the diquark current as 
$\overline{\psi}i\gamma_{5}\tau_{2}J_{a}\psi_{C} = (\Delta_a + \delta _a)/(2 G_D)$ 
and drop the quadratic term in $\delta _a$, where the diquark condensate is 
$\Delta_{a}=2G_{D}\left\langle
\overline{\psi}i\gamma_{5}\tau_{2}J_{a}\psi_{C}\right\rangle $.
The diquark condensate fluctuation can be introduced by the replacement 
$\Delta_{a}\rightarrow \Delta_{a}+\varphi _a$ and keeping quadratic terms 
in the fluctuation $\varphi_a$. The above operation is equivalent to 
performing the Hubbard-Stratonovich transformation in the diquark sector. 
The Lagrangian (\ref{eq:njl01}) now becomes
\begin{eqnarray}
\mathcal{L}_{NJL} & \approx & -\frac{1}{2}\overline{\Psi}S^{-1}\Psi
-\frac{1}{4G_{D}}\sum_{a}|\Delta_{a}|^{2}
-G_{S}(\sigma_{u}+\sigma_{d})^2 \nonumber \\
 &  & -\frac{1}{8G_{D}}(\varphi_{aR}^{2}+\varphi_{aI}^{2})
+\frac{1}{2}\overline{\Psi}\varphi_{ai}\widehat{\Gamma}_{ai}\Psi\;.
\label{eq:njl04}
\end{eqnarray}
Here $\Psi=(\psi,\psi_{C})^{T}$ and 
$\overline{\Psi}=(\overline{\psi},\overline{\psi}_{C})$
are quark spinors in the Nambu-Gorkov (NG) basis. The charge-conjugate
spinors are defined by $\psi_{C}=C\overline{\psi}^{T}$ and
$\overline{\psi}_{C}=\psi^{T}C$ with $C=i\gamma^{2}\gamma^{0}$.
The complex diquark fluctuation $\varphi_a$ 
has been decomposed in terms of its real and imaginary
parts, $\varphi_a=(\varphi_{aR}+i\varphi_{aI})/\sqrt{2}$,  
with color indices $a=1,2,3$. 
The inverse fermion propagator $S^{-1}$ in the NG basis is given by
\begin{equation}
S^{-1}(P)=-\left(\begin{array}{cc}
P_{\mu}\gamma^{\mu}+\hat{\mu}\gamma^{0}
-\hat{m} & i\gamma_{5}\tau_{2}J_{a}\Delta_{a}^{\dagger}\\
i\gamma_{5}\tau_{2}J_{a}\Delta_{a} &
P_{\mu}\gamma^{\mu}-\hat{\mu}\gamma^{0}
-\hat{m}\end{array}\right).
\label{eq:fermion-prop}
\end{equation}
where $\hat{m}=\mathrm{diag}(m_u,m_d)$ is the quark mass matrix with
corrections from chiral condensates, 
$m_i=m_i^{(0)}-2G_S(\sigma_{u}+\sigma_{d})$ with $i=u,d$.
The quark-quark-diquark vertices $\widehat{\Gamma}_{ai}$
are given by
$\widehat{\Gamma}_{aR}=\frac{i}{\sqrt{2}}\gamma_{5}\tau_{2}J_{a}\tau_{1}^{NG}$,
$\widehat{\Gamma}_{aI}=\frac{i}{\sqrt{2}}\gamma_{5}\tau_{2}J_{a}\tau_{2}^{NG}$,
where $\tau_{s}^{NG}$ ($s=1,2,3$) are Pauli matrices in NG space.
In the following, without loss
of generality we choose the diquark condensate
to be $\Delta_{a}=\delta_{a3}\Delta_{3}$.
Note that we only consider the scalar channel for the diquark
condensate, as we are only interested
in the lowest baryon state, not the higher-lying excited ones.
Including the axial-vector channel is straightforward,
but will not modify our results qualitatively.
Finally, we remark that the tadpole term 
$\varphi _a \Delta _a^*+ \varphi _a^*\Delta _a$,
which in principle also appears in Eq.\ (\ref{eq:njl04}),
is cancelled by the term 
$\varphi _a \overline{\psi}_C i\gamma_{5}\tau_{2}J_{a}\psi + 
\varphi _a^* \overline{\psi} i\gamma_{5}\tau_{2}J_{a}\psi _C$ 
at the one-loop level, where $\overline{\psi}_C \psi + \overline{\psi}\psi _C$ 
contracts and forms a quark loop in the NG basis. 
The cancellation condition is just the gap equation for $\Delta$.

We now add the baryon field to our Lagrangian. We assume
the baryon to be generated by an interaction term between two quark and
two diquark fields,
\begin{eqnarray}
{\cal L}_{B} & = & G_B \varphi_a^\dagger \bar{\psi}_a \psi_b \varphi_b
\nonumber \\
& \simeq & -\frac{1}{2G_{B}}\overline{\mathbf{B}}\mathbf{B}
+\frac 12\overline{\mathbf{B}}\widehat{\Gamma}_{Bi}\Psi_{a}\varphi_{ai}
+\frac 12\varphi_{ai}\overline{\Psi}_{a}\widehat{\Gamma}_{Bi}^{*}\mathbf{B}\;.
\label{eq:B}
\end{eqnarray}
Here, we decomposed 
$\psi_a \varphi_a = \left\langle \psi_a \varphi_a \right\rangle +
\beta_a$, defined the baryonic field as 
$B =G_B \left\langle \psi_a \varphi_a \right\rangle$, 
and neglected terms of order $O(\beta_a^2)$. 
The baryonic fields in the NG basis are then
denoted by $\mathbf{B}=(B,B_{c})^{T}$
and $\overline{\mathbf{B}}=(\overline{B},\overline{B}_{c})$.
The baryon-quark-diquark vertices are
$\widehat{\Gamma}_{BR}=\frac{1}{\sqrt{2}}1_{NG}$
and $\widehat{\Gamma}_{BI}=i\frac{1}{\sqrt{2}}\tau_{3}^{NG}$, respectively.
The sum of the Lagrangians (\ref{eq:njl04}) and (\ref{eq:B}) is the starting point
for our further treatment. In the following, for the sake of simplicity
we assume exact isospin symmetry and we work in the chiral limit,
i.e., $\sigma_{u}=\sigma_{d}\equiv\sigma$, thus $m_u=m_d=m_q$,
$\mu_{u}=\mu_{d}=\mu_{q}$, and $m_{u}^{(0)}=m_{d}^{(0)}\equiv 0$.

We now derive the full diquark propagator via
the Dyson-Schwinger equation,
\begin{eqnarray}
D_{i,a}^{-1}(p_{0},\mathbf{p}) & = &
-\frac{1}{4G_{D}}-\Pi_{i,a}(p_{0},\mathbf{p})\;,
\end{eqnarray}
where $p_{0}=i2\pi nT$are the bosonic Matsubara
frequencies ($n =0,\pm 1,\pm 2, \ldots$), $i,j=R,I$,
and $a,b=1,2,3$ are fundamental colors.
The full propagator $D_{i,a}$ and the self-energy $\Pi_{i,a}$ only
carry one index $i=R,I$ and one color index $a$, because they are diagonal
in the space of $R,I$ and in color space.
The self-energy has the property
$ \Pi_{R/I,a}  = \frac{1}{2} \left( \Pi_{0}^{a}\pm
  \Pi_{1}^{a}\right)$,
where $\Pi_{0}^{a}$ and $\Pi_{1}^{a}$ depend on the diagonal and the off-diagonal
parts of the quark propagator, respectively, and $\Pi_{1}^{a}=\delta_{a3}\Pi_{1}^{3}$.
The expressions for $\Pi_{0}^{a}(p_0,\mathbf{p})$ and $\Pi_{1}^{a}(p_0,\mathbf{p})$ are 
\begin{eqnarray}
\Pi_{0}^{1,2}&=&2\int\frac{d^{3}k}{(2\pi)^{3}} \, c_{k,p+k}\nonumber \\
&& \times \left[ \frac{e_{1}'\epsilon_{k}^{e'}+\xi_{k}^{e'}}{2e_{1}'\epsilon_{k}^{e'}}
\frac{1-f(e_{1}'\epsilon_{k}^{e'})-f(\xi_{p+k}^{e})}{p_{0}-e_{1}'\epsilon_{k}^{e'}-\xi_{p+k}^{e}} \right. \nonumber\\
&&+\left. \frac{e_{1}\epsilon_{p+k}^{e}+\xi_{p+k}^{e}}{2e_{1}\epsilon_{p+k}^{e}}
\frac{1-f(\xi_{k}^{e'})-f(e_{1}\epsilon_{p+k}^{e})}{p_{0}-\xi_{k}^{e'}-e_{1}\epsilon_{p+k}^{e}}\right]\;, \nonumber\\
\Pi_{0}^{3}&=&4\int\frac{d^{3}k}{(2\pi)^{3}}\frac{e_{1}'\epsilon_{k}^{e'}+\xi_{k}^{e'}}{2e_{1}'\epsilon_{k}^{e'}}
\frac{e_{1}\epsilon_{p+k}^{e}+\xi_{p+k}^{e}}{2e_{1}\epsilon_{p+k}^{e}}\nonumber
\\&&\times \frac{1-f(e_{1}'\epsilon_{k}^{e'})-f(e_{1}\epsilon_{p+k}^{e})}{p_{0}-e_{1}'\epsilon_{k}^{e'}-e_{1}
\epsilon_{p+k}^{e}}c_{k,p+k}, \nonumber\\
\Pi_{1}^{1,2}&=&0,\nonumber\\ 
\Pi_{1}^{3} &=&-\int\frac{d^{3}k}{(2\pi)^{3}}\frac{\Delta_{3}^{2}}{e_{1}e_{1}'\epsilon_{k}^{e}\epsilon_{p+k}^{e'}}\nonumber\\
&&\times\frac{1-f(e_{1}\epsilon_{k}^{e})-f(e_{1}'\epsilon_{p+k}^{e'})}{p_{0}-e_{1}\epsilon_{k}^{e}-e_{1}'\epsilon_{p+k}^{e'}}c_{k,p+k},
\end{eqnarray}
where summations over $e,e',e_1,e_1'=\pm 1$ are implied, 
$f(x)=1/(e^{x/T}+1)$ is the Fermi-Dirac distribution, 
$E_{k}=\sqrt{k^{2}+m_{q}^{2}}$, $\xi_{k}^{e}=eE_{k}-\mu$, 
$\epsilon_{k}^{e}=\sqrt{(\xi_{k}^{e})^{2}+\Delta^{2}}$, 
and $c_{k,p+k}= 1+ee'\frac{\mathbf{k}\cdot(\mathbf{p+k})+m_q^2}{E_{k}E_{p+k}}$.

Some simple properties of $D_{i,a}^{-1}$ are: (1)
$D_{R,3}^{-1}\neq D_{I,3}^{-1}$ when $\Delta_{3}\neq0$; (2)
$D_{i,1}^{-1}=D_{j,2}^{-1}$ for any $i,j=R,I$; (3)
$D_{i,1}^{-1}=D_{i,2}^{-1}=D_{i,3}^{-1}=D^{-1}$ when $\Delta_{3}=0$
for any $i=R,I$. We also have $\Pi_{1}^a=0$ when $\Delta_{a}=0$. The
spectral density for diquarks is then given by
\begin{eqnarray}
\lefteqn{\rho_{i,a}(\omega,\mathbf{p})= \frac{1}{\pi} } \nonumber \\
& \times & \frac{\mathrm{Im}D_{i,a}^{-1}(\omega+i\eta,\mathbf{p})}{
\left[\mathrm{Re}D_{i,a}^{-1}(\omega+i\eta,\mathbf{p})\right]^{2}
+\left[\mathrm{Im}D_{i,a}^{-1}(\omega+i\eta,\mathbf{p})\right]^{2}}\;,
\label{eq:spectral-density}
\end{eqnarray}
where we analytically continued $p_{0}\rightarrow\omega+i\eta$
with real $\omega$ and $\eta$ a small positive number. We have similar properties for
the spectral densities as for $D_{i,a}^{-1}$. With the spectral density,
we can obtain the full propagator via the dispersion relation
\begin{equation}
D_{i,a}(p_{0},\mathbf{p})=\int_{-\infty}^{\infty}d\omega
\frac{\rho_{i,a}(\omega,\mathbf{p})}{\omega-p_{0}}\;.
\label{eq:disp-full}
\end{equation}

From the Lagrangian (\ref{eq:B}) the 11-component in NG space of the inverse baryon
propagator is $ S_{B}^{-1}=-1/(2G_{B})-\Sigma$, where
\begin{equation}
\Sigma(P)=-\frac 14 \sum_{a}\int_{K}S_{11}^{a}(P-K)[D_{R,a}(K)+D_{I,a}(K)]
\label{eq:baryon-selfen}
\end{equation}
is the 11-component of the baryon self-energy.
The quark propagator in NG space, $S_{11}^a$, is diagonal in color
space. In the presence of a non-vanishing diquark condensate,
$S_{11}^{1}=S_{11}^{2}\neq S_{11}^{3}$. If the diquark condensate vanishes,
$S_{11}^{1}=S_{11}^{2}=S_{11}^{3}$ and $D_{R,a}=D_{I,b}$ for any
$a,b$. In order to evaluate $\Sigma$, we insert
Eq.\ (\ref{eq:disp-full}) into Eq.\ (\ref{eq:baryon-selfen}).
Since we are interested in baryons at rest, we shall take the $\mathbf{p}= \mathbf{0}$
limit of the positive energy component of $S_{B}^{-1}$,
$S_{B,+}^{-1}(p_{0},\mathbf{p}=\mathbf{0})=
\frac{1}{2}\mathrm{Tr}\left[S_{B}^{-1}\Lambda_{\mathbf{p=0}}^{+}\gamma^{0}\right]$,
where $\Lambda_{\mathbf{p}}^{s}$ is the energy projector
$\Lambda_{\mathbf{p}}^{s}=\frac{1}{2}\left[1+s \left(
    \gamma_{0}\gamma\cdot\mathbf{p} +\gamma_{0}M_{B}\right)/E_{p} \right]$,
with $E_{p}=\sqrt{p^{2}+M_{B}^{2}}$ and $s=\pm 1$. In the
homogeneous limit, $\mathbf{p}= \mathbf{0}$,
the energy projector assumes a simple form,
$\Lambda_{\mathbf{p=0}}^{s}=\frac{1}{2}(1+s\gamma_{0})$,
which is independent of $M_{B}$. Then, we obtain the spectral density as
\begin{eqnarray}
\lefteqn{\rho_{B}(\omega,\mathbf{p})=\frac{1}{\pi}}\nonumber \\
& \times & \frac{\mathrm{Im}S_{B,+}^{-1}(\omega+i\eta,\mathbf{0})}{
\left[\mathrm{Re}S_{B,+}^{-1}(\omega+i\eta,\mathbf{0})\right]^{2}
+\left[\mathrm{Im}S_{B,+}^{-1}(\omega+i\eta,\mathbf{0})\right]^{2}}\;,
\label{eq:baryon-spec-density}
\end{eqnarray}
where we have again analytically continued $p_{0}\rightarrow\omega+i\eta$.

\begin{figure}
\includegraphics[scale=0.5]{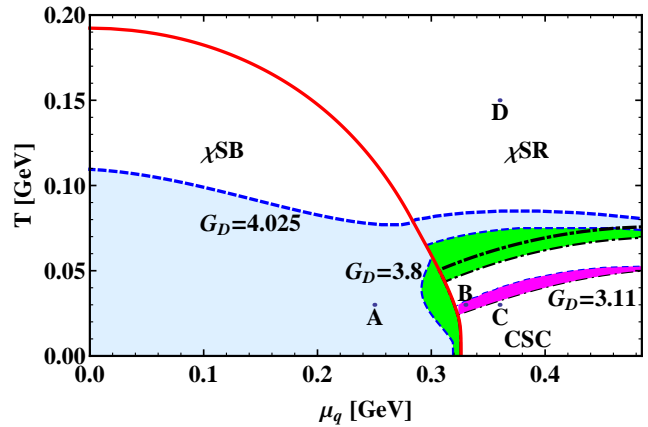}

\caption{\label{fig:diss-diquark}(color online) The phase
diagram obtained within our model. For explanations see text.}

\end{figure}

\begin{figure}

\includegraphics[scale=0.4]{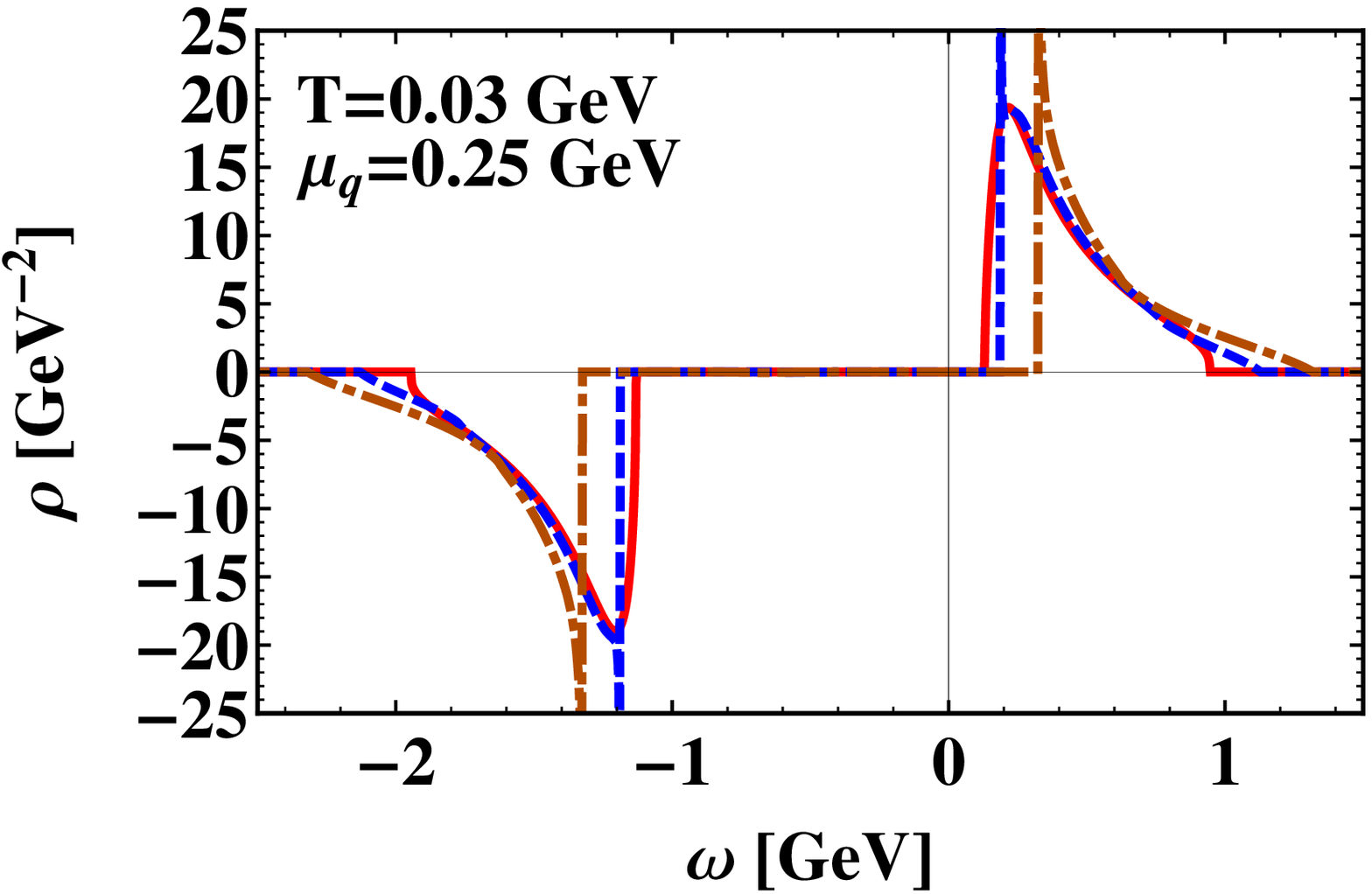}
\includegraphics[scale=0.4]{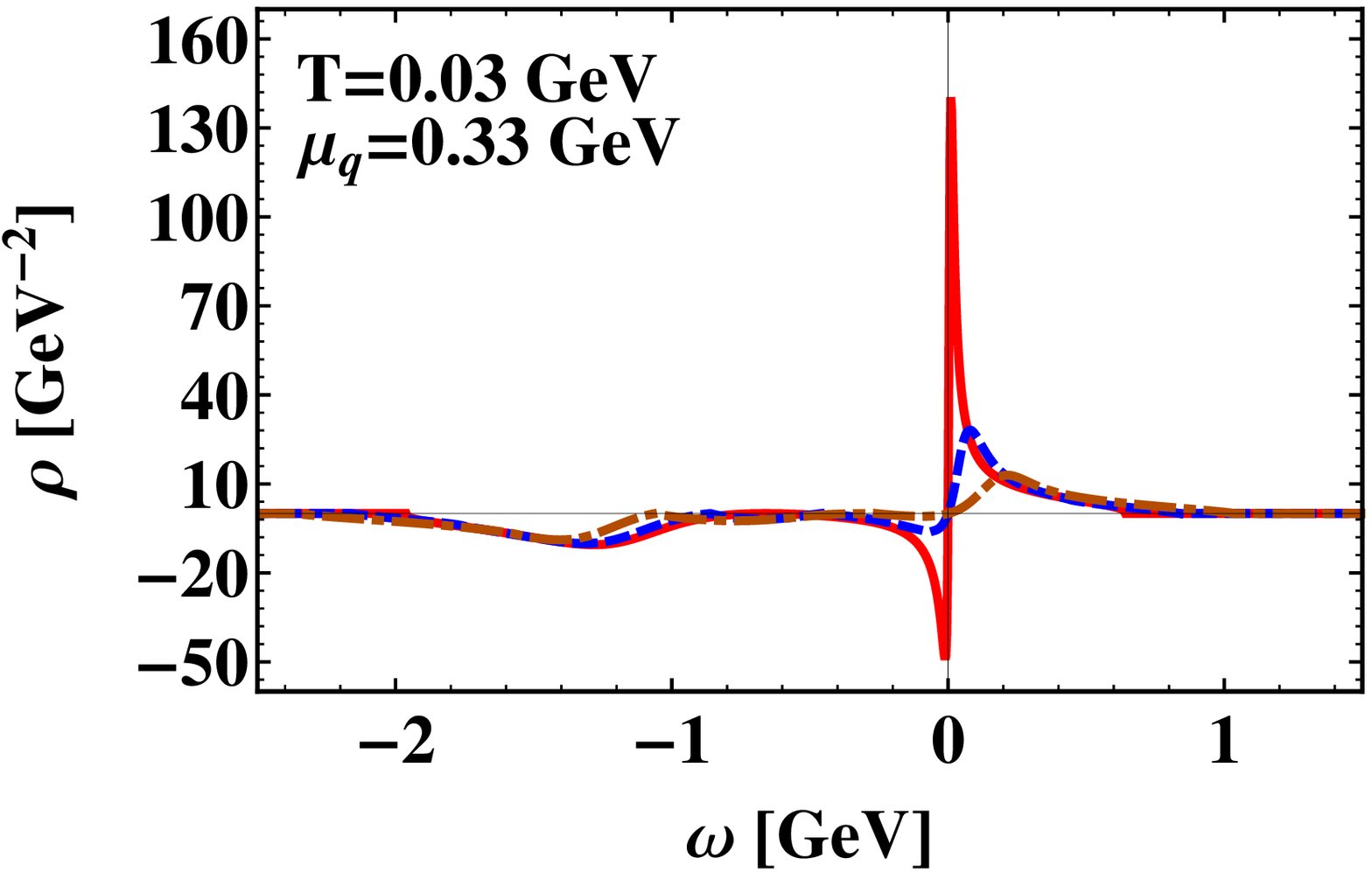}
\includegraphics[scale=0.4]{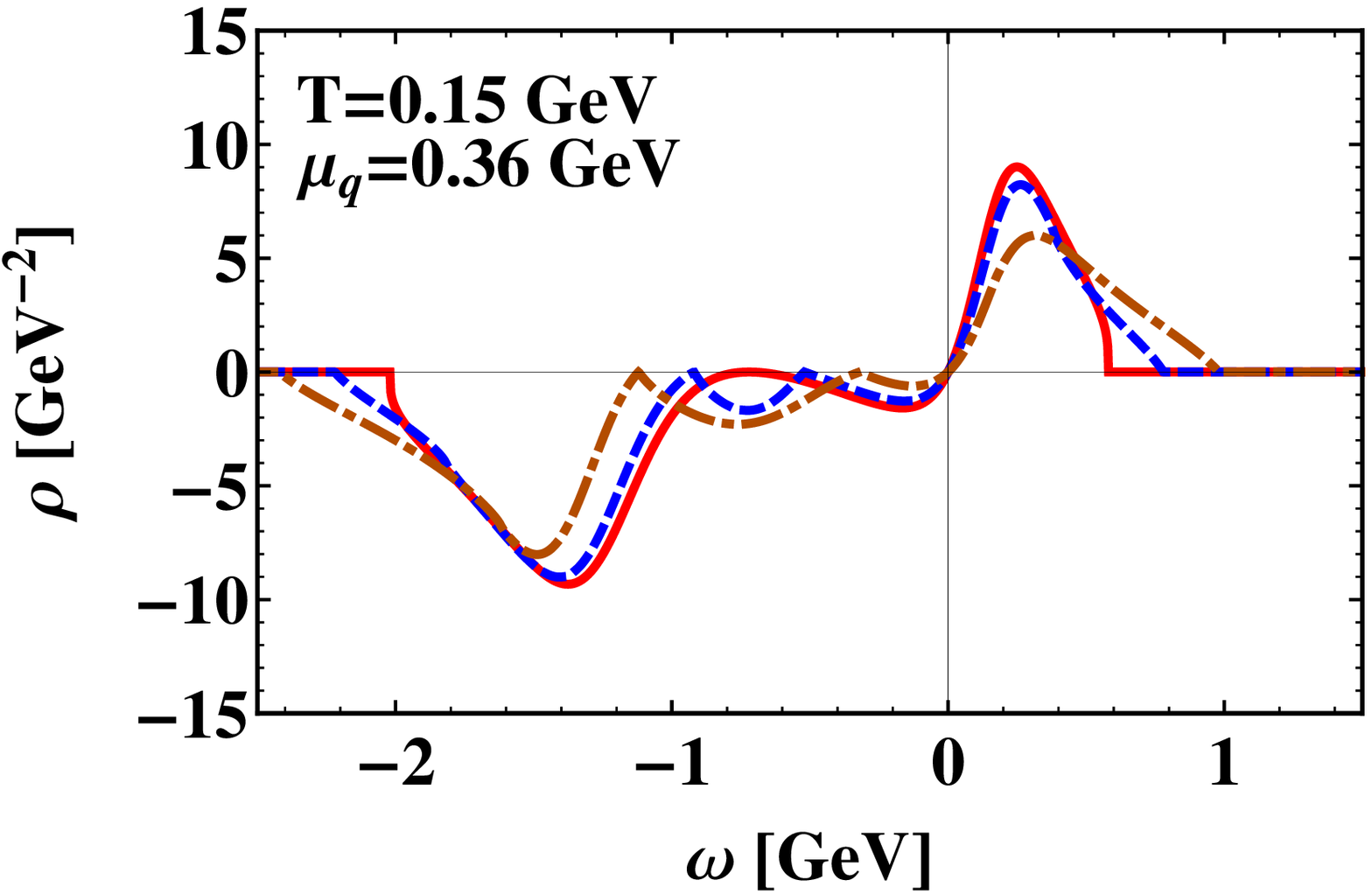}

\caption{\label{fig:diquark-spectral-density}(color online) Diquark spectral
densities for different values of $T$ and $\mu_{q}$. 
The upper panel corresponds to the point A in the phase diagram with $(T,\mu_{q})=(0.03,0.25)$.
The middle panel corresponds to the point B in the phase diagram
with $(T,\mu_{q})=(0.03,0.33)$. 
The lower panel is in the chiral
symmetric phase, corresponding to the point D in the phase
diagram with $(T,\mu_{q})=(0.15,0.36)$. For all panels we have $\rho\equiv\rho_{R}=\rho_{I}$. 
The red solid lines are for $p=0$,
the blue dashed and brown dash-dotted lines are for
$p=0.2$ and $p=0.4$, respectively. All units in GeV.} 
\end{figure}

\begin{figure}

\includegraphics[scale=0.4]{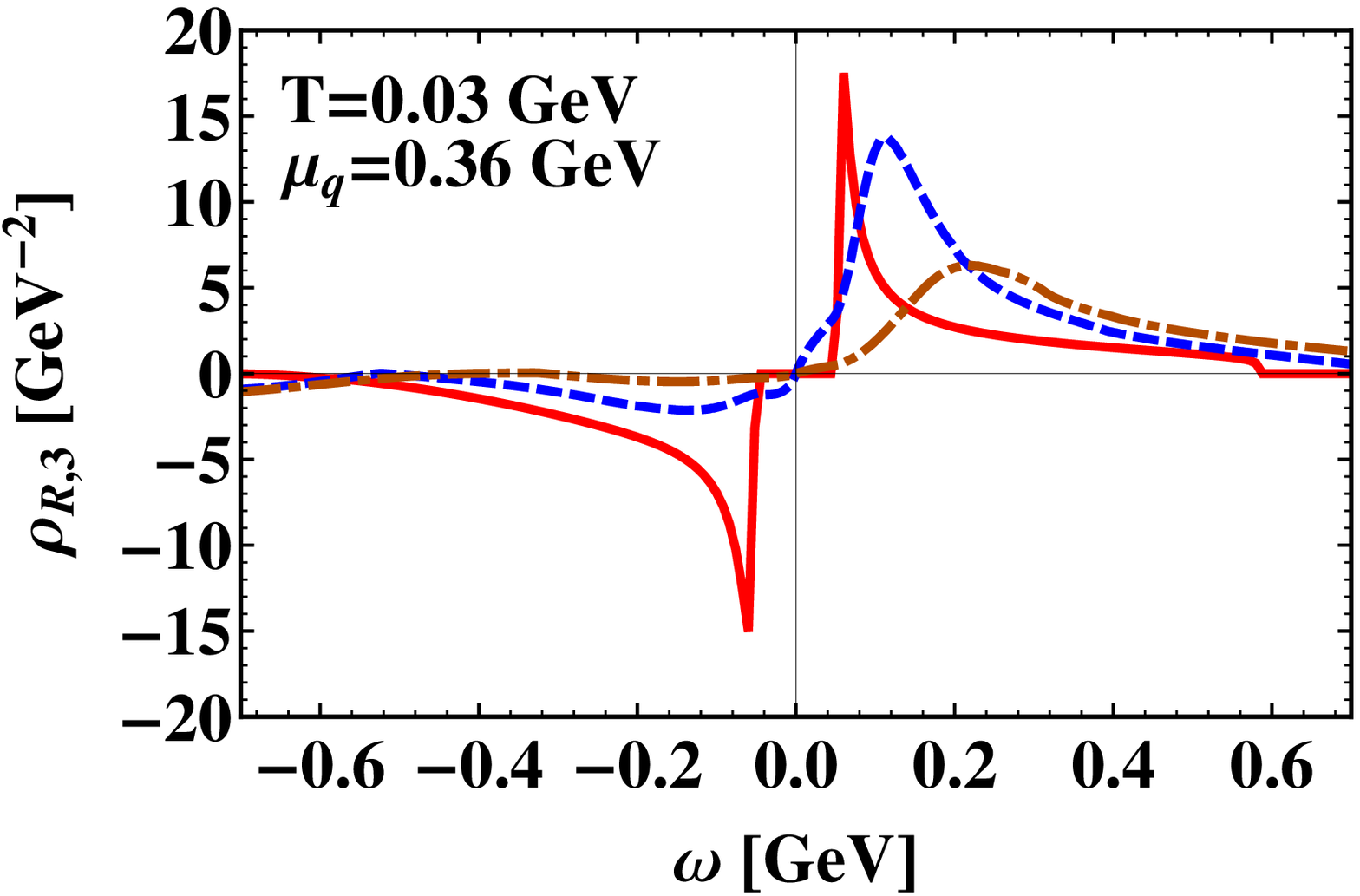}
\includegraphics[scale=0.4]{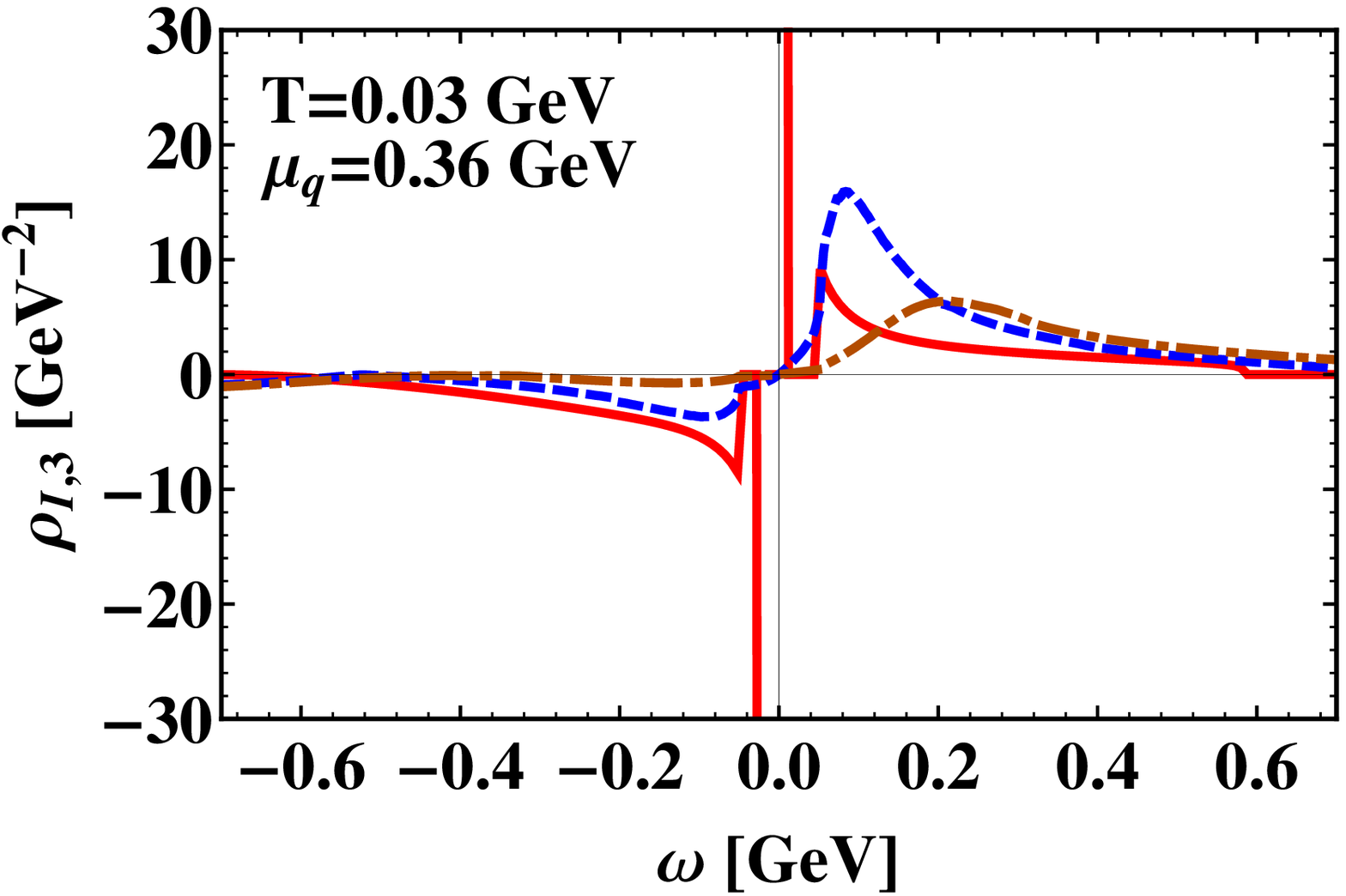}
\includegraphics[scale=0.4]{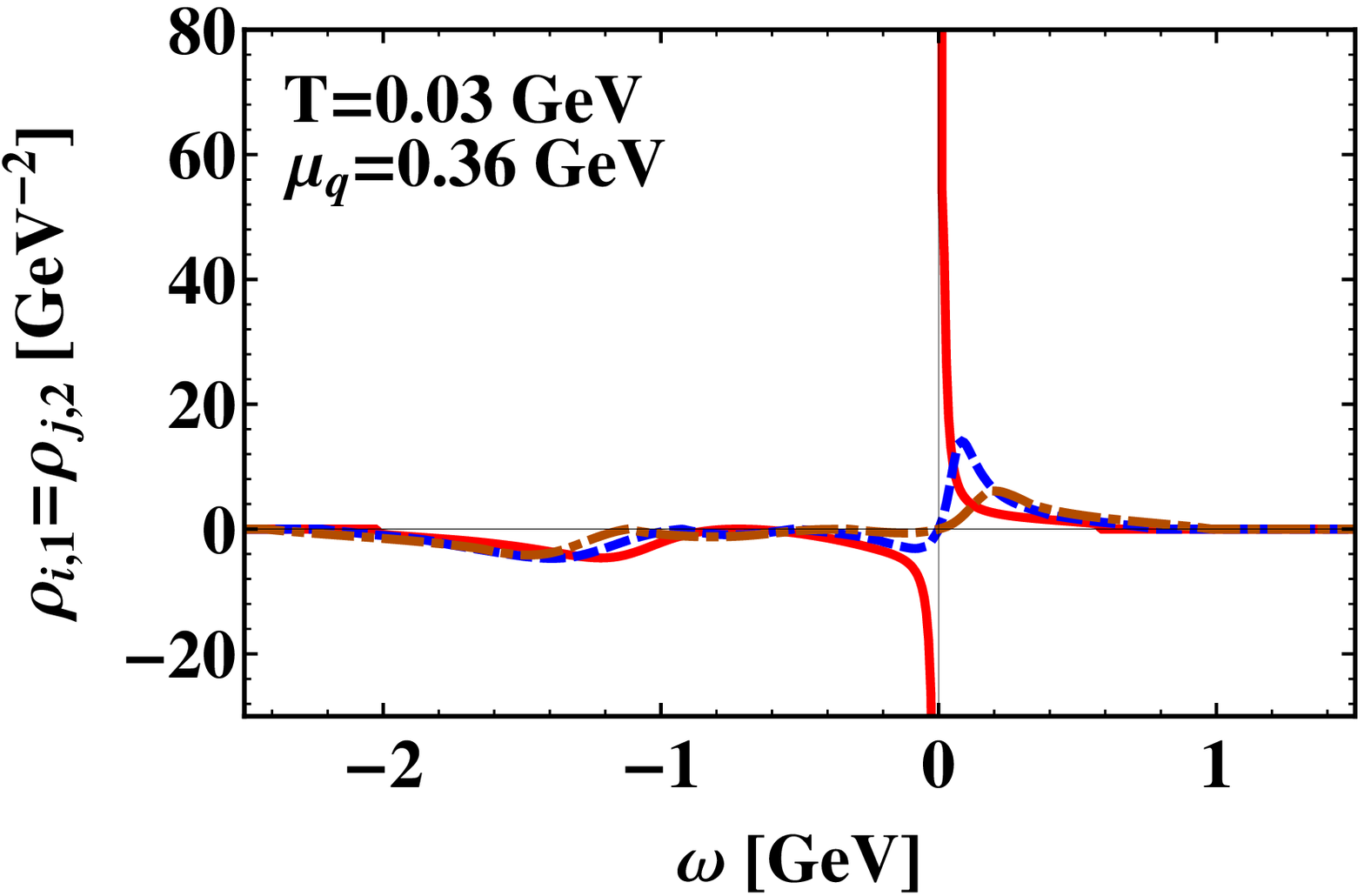}

\caption{\label{fig:diquark-spectral-density1}(color online) Diquark spectral
densities for different values of $T$ and $\mu_{q}$. 
The upper, middle and lower panels are $\rho_{R,3}$, $\rho_{I,3}$ 
and $\rho_{I/R,1/2}$ in the CSC phase, 
respectively, corresponding to the point C in the phase diagram with
$(T,\mu_{q})=(0.03,0.36)$. The red solid lines are for $p=0$,
the blue dashed and brown dash-dotted lines are for
$p=0.2$ and $p=0.4$, respectively. All units in GeV. }

\end{figure}

\begin{figure}
\includegraphics[scale=0.4]{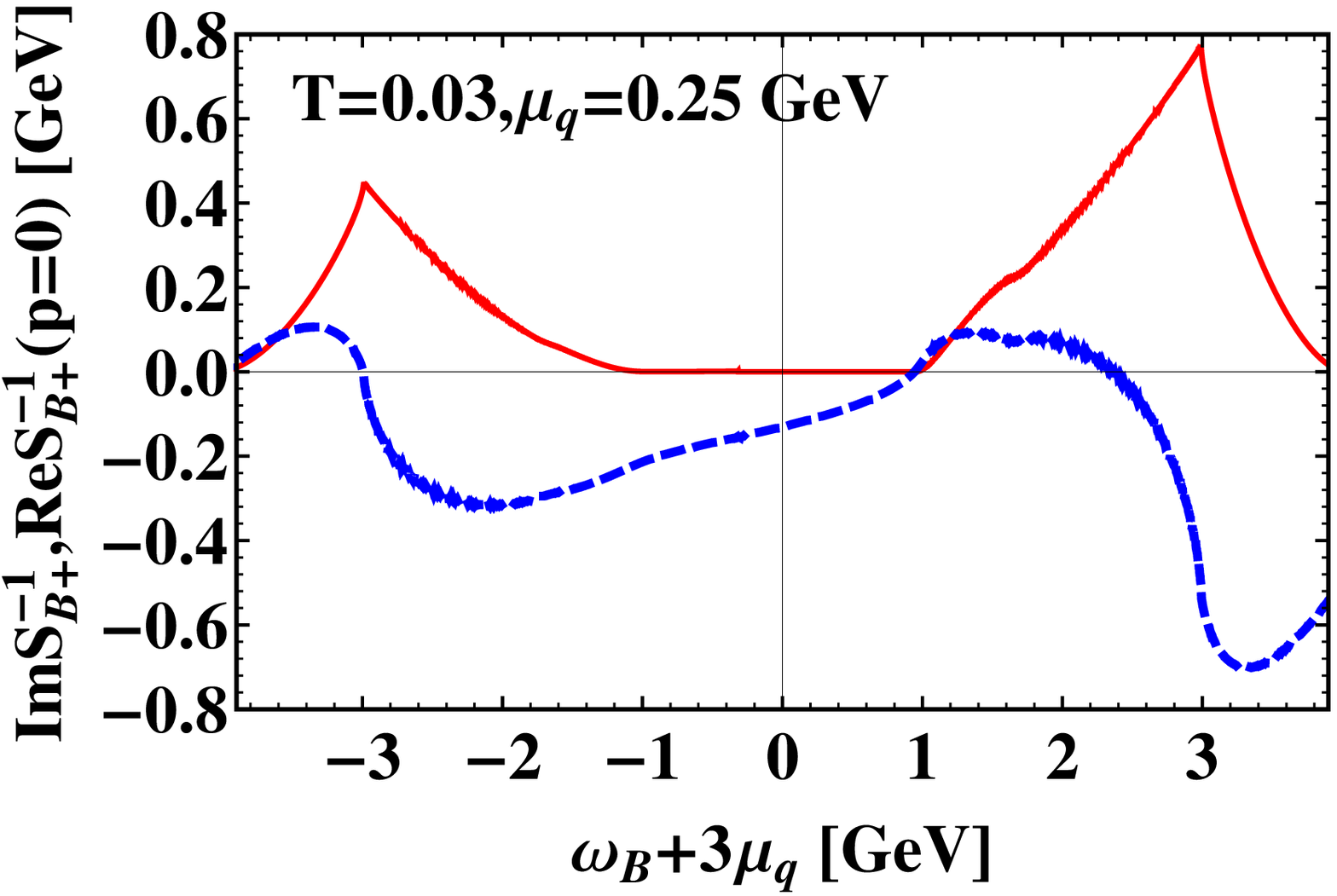}
\includegraphics[scale=0.4]{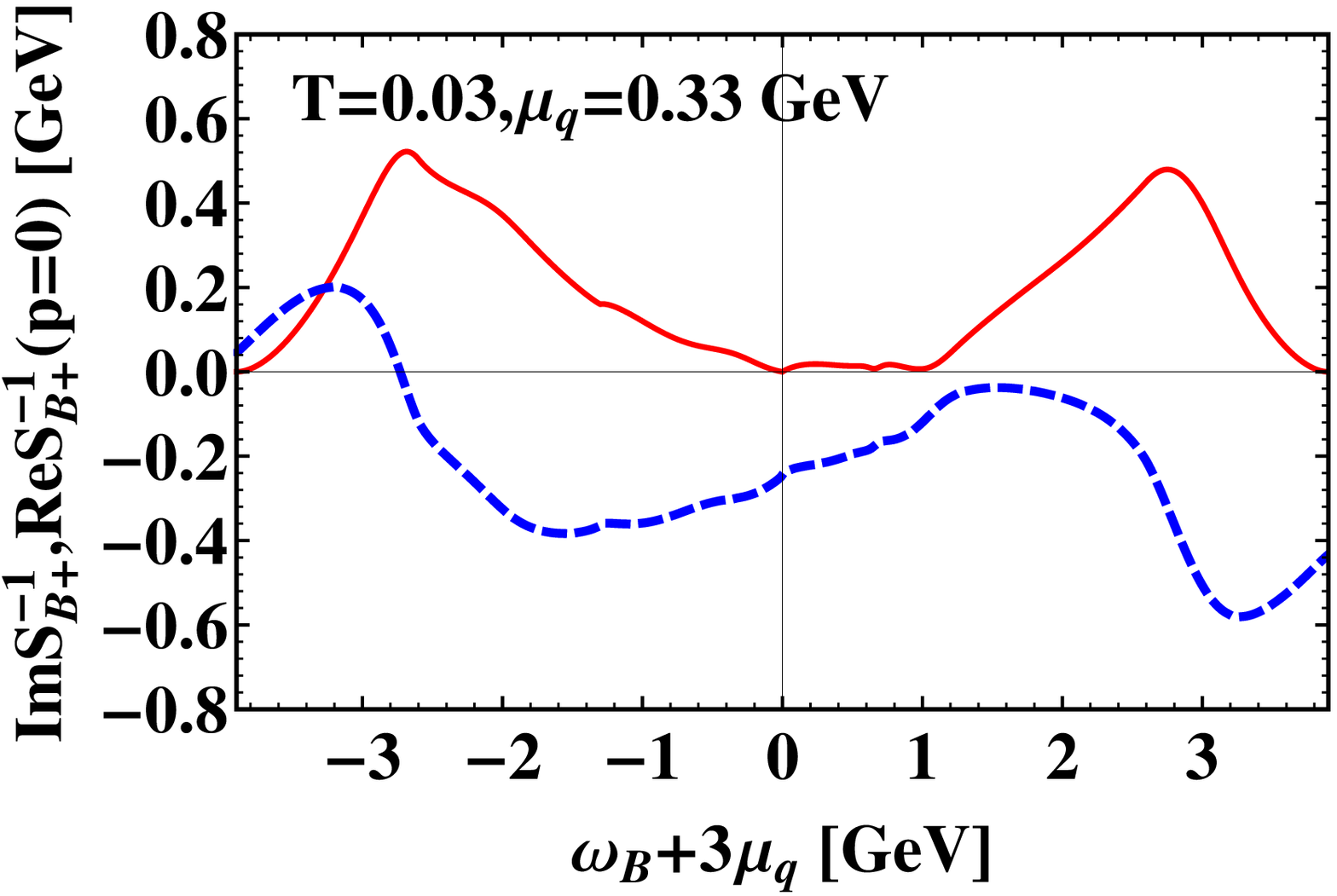}
\includegraphics[scale=0.4]{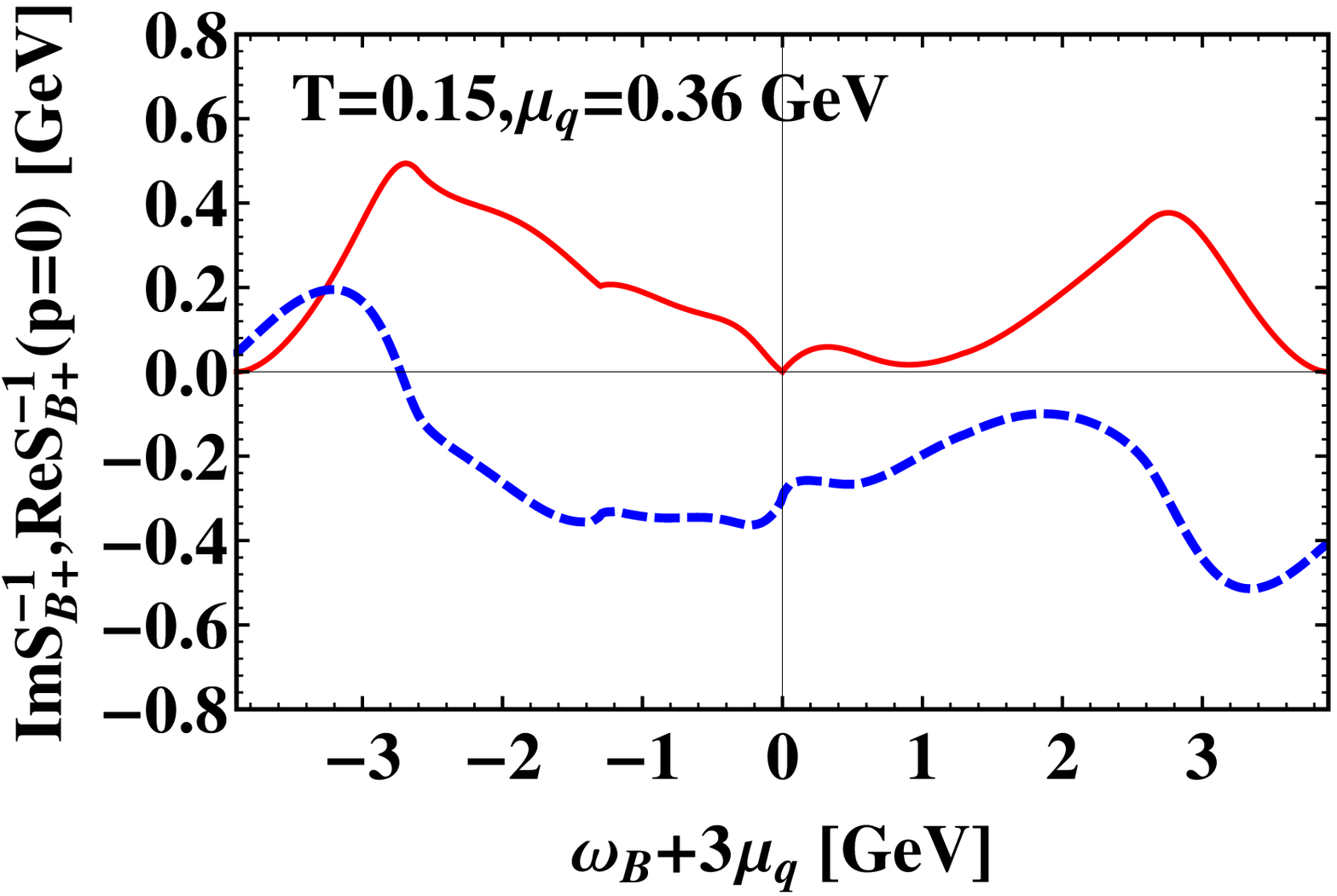}
\includegraphics[scale=0.4]{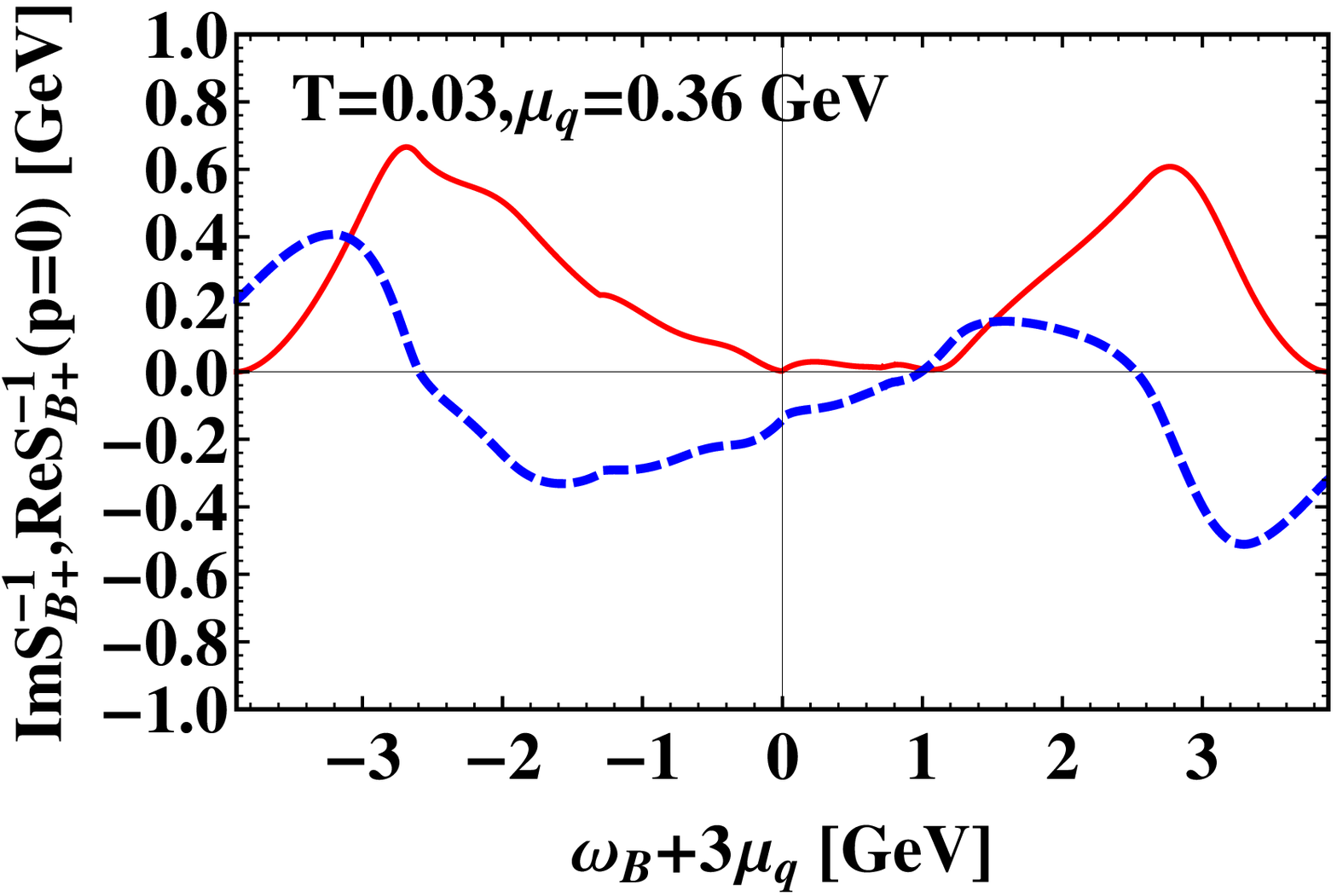}

\caption{\label{fig:re-im-pole-l}(color online) The real (blue dashed) and
imaginary (red solid) parts of the inverse propagators for baryons as
functions of energy $\omega$ at different $T$ and $\mu_{q}$.
From top to bottom, the first panel: $T=0.03$ and $\mu_{q}=0.25$ (point A). 
The second panel: $T=0.03$ and $\mu_{q}=0.33$ (point B). The third panel: $T=0.15$ and
$\mu_{q}=0.36$ (point D). The fourth panel: $T=0.03$ 
and $\mu_{q}=0.36$ (point C). All units in GeV. }

\end{figure}

\begin{figure}

\includegraphics[scale=0.4]{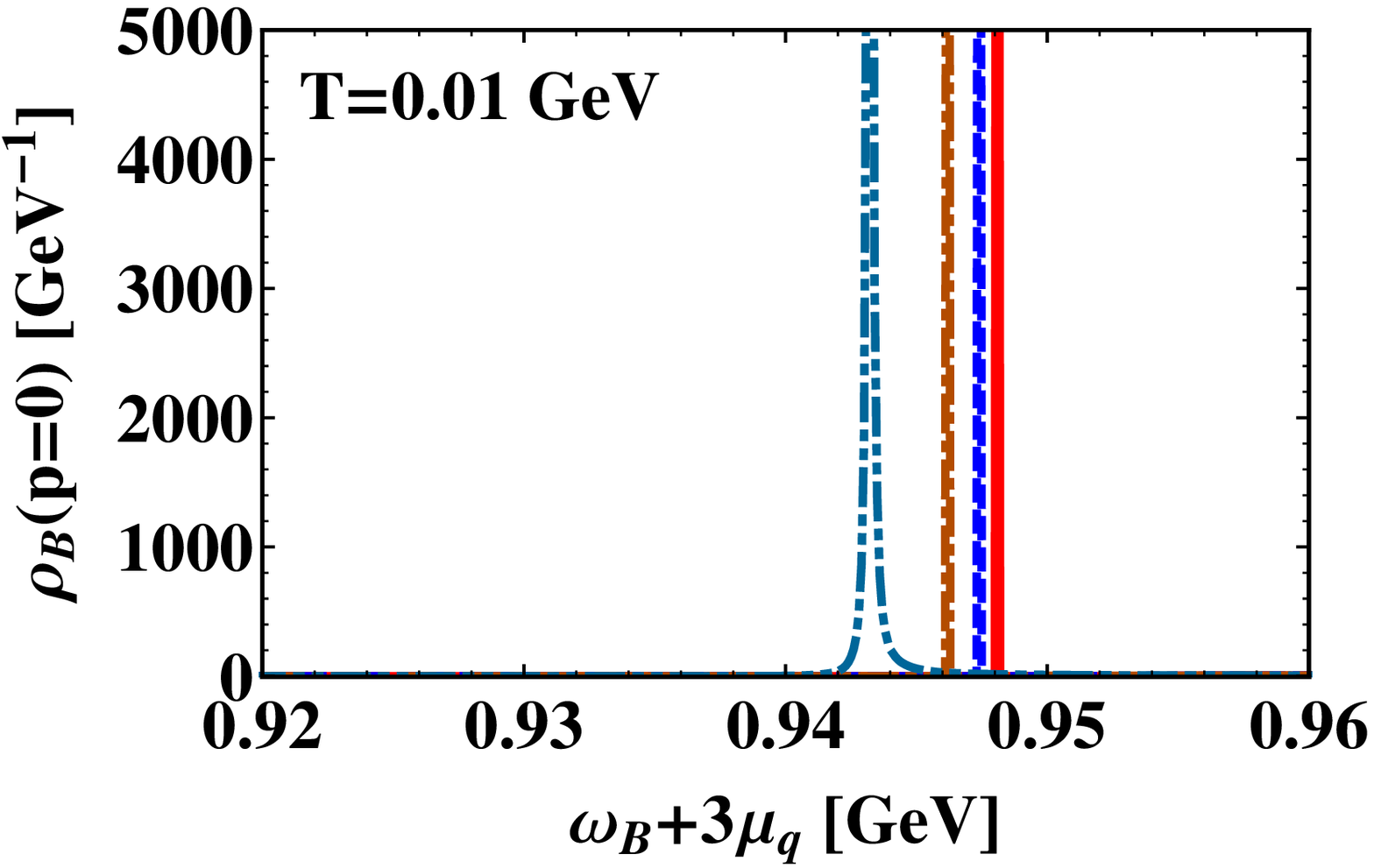}
\includegraphics[scale=0.4]{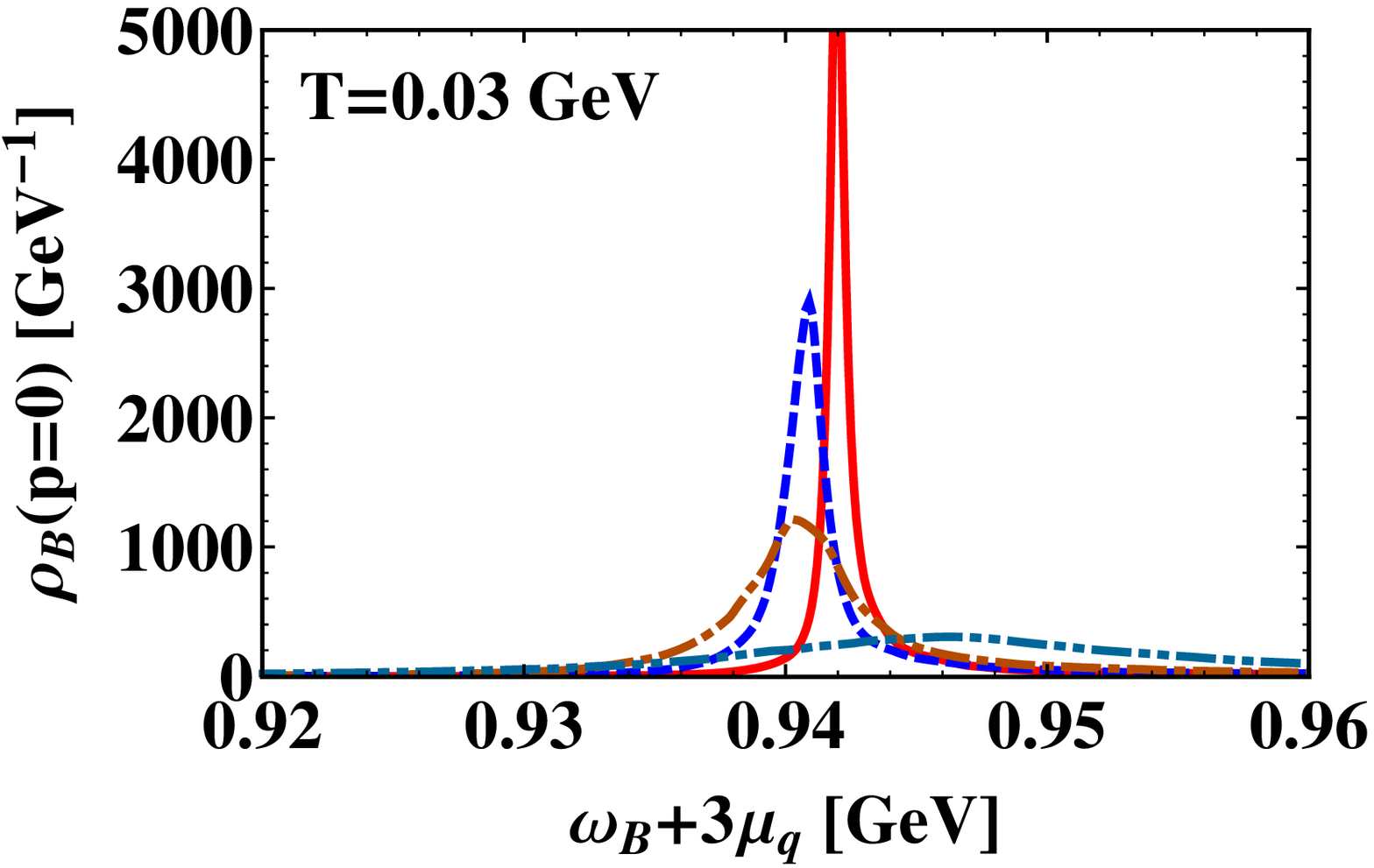}
\includegraphics[scale=0.4]{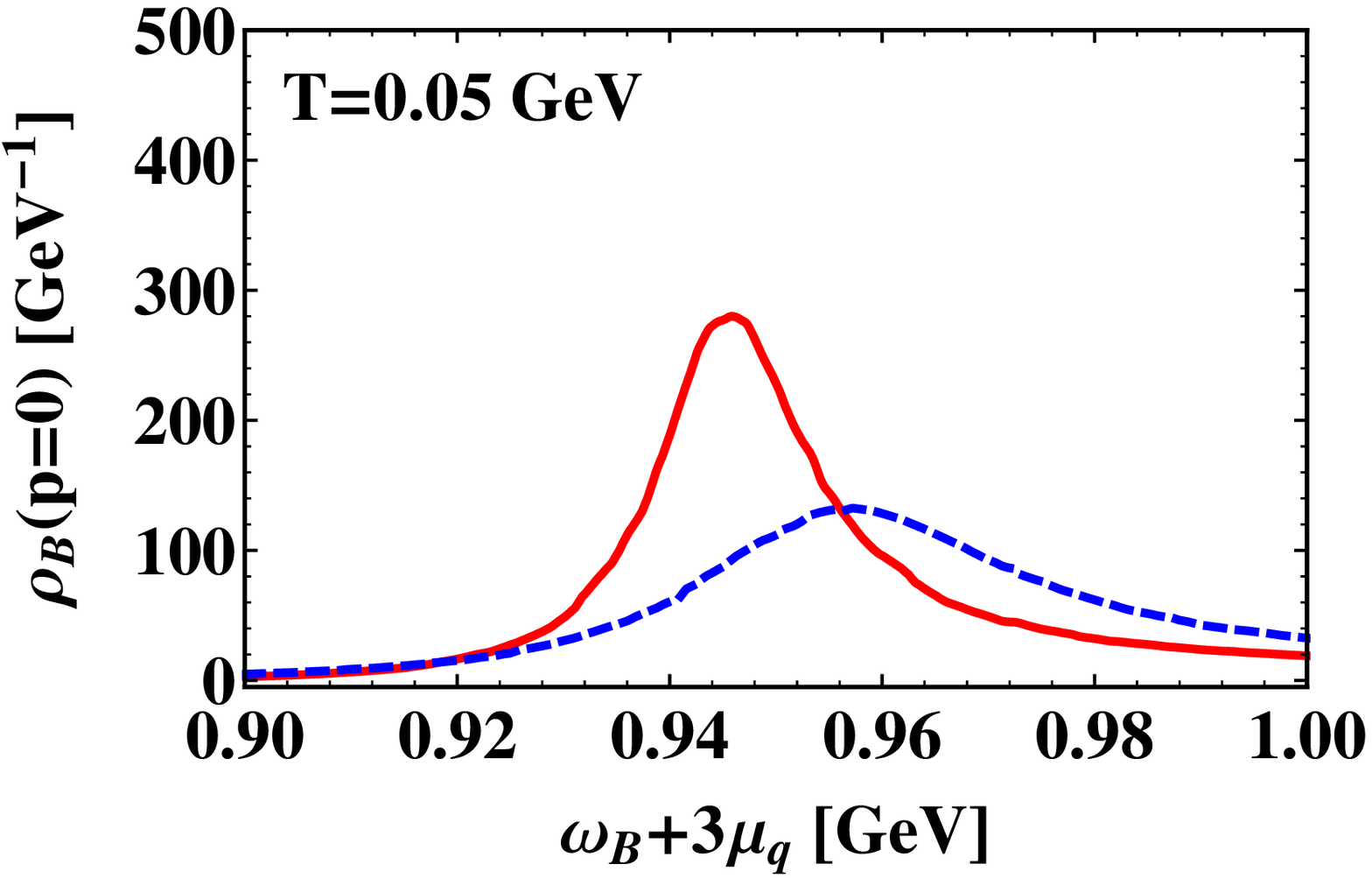}
\includegraphics[scale=0.4]{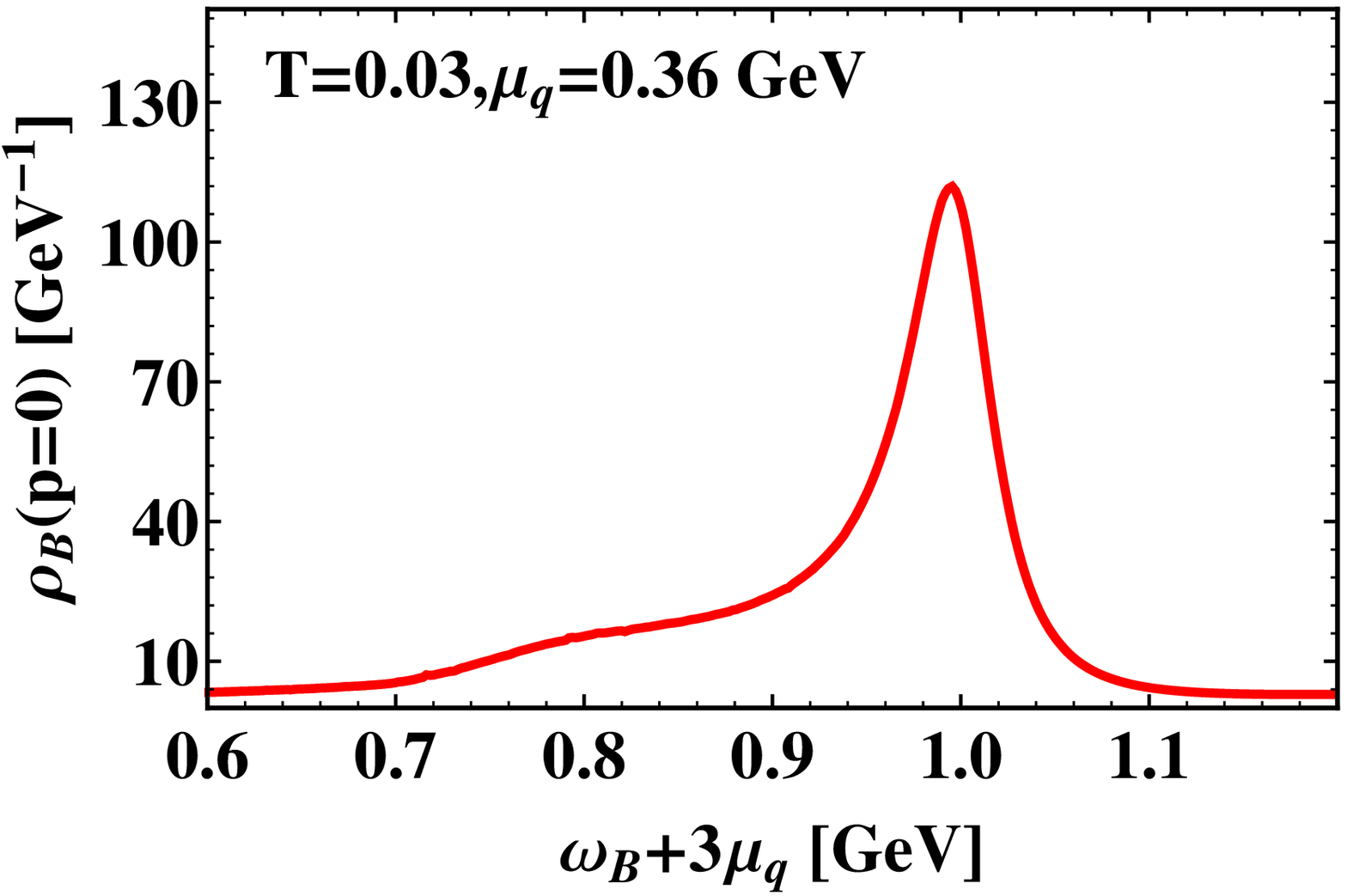}

\caption{\label{fig:baryon-spec}(color online) The baryon spectral
densities at different values of $T$ and $\mu_{q}$ as functions of
$\omega+3\mu_{q}$ for $\mathbf{p}=\mathbf{0}$. Four values of
$\mu_{q}$ are chosen for each panel, 0.29 GeV (red solid), 0.30 GeV
(blue dashed), 0.31 GeV (brown dash-dotted) and 0.32 GeV (light blue
dash-dot-dotted). In the fourth panel (from top to bottom) we show the result for
$T=0.03$ GeV and $\mu_{q}=0.36$ GeV (point C of Fig.\ 
\ref{fig:diss-diquark}).}
\end{figure}

\begin{figure}

\includegraphics[scale=0.5]{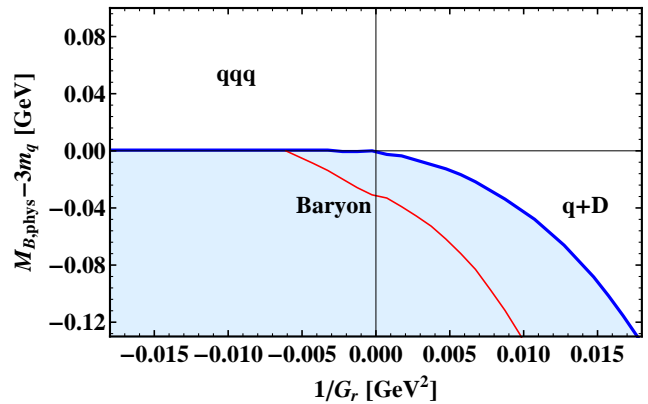}

\caption{(color online)
The quantity $\omega_B+3(\mu_q - m_q)$ as a function
of the inverse renormalized coupling $1/G_r$ at $T=\mu_{q}=0$.
In the shaded region, i.e., below the blue curve, baryons are stable.
The red curve inside this region is given by $M_{B,phys}-3 m_q$,
where $M_{B,phys}$ is the physical mass of the baryon. }
\label{fig:baryonstate}

\end{figure}

In our calculations for Figs.\ \ref{fig:diss-diquark}--\ref{fig:baryonstate}, 
we choose the following parameters: $G_{S}=5.1\;\mathrm{GeV}^{-2}$, 
$\Lambda=0.65\;\mathrm{GeV}$ (momentum cutoff). 
For Figs.\ \ref{fig:diss-diquark} and \ref{fig:baryonstate}, we vary $G_D$, 
in order to investigate the effect of the diquark coupling constant on 
the boundaries of the diquark dissociation and the color-superconducting (CSC) phase 
and on the baryon formation. 
For Figs.\ \ref{fig:diquark-spectral-density}--\ref{fig:baryon-spec}, 
we set $G_D = 3.11\; \mathrm{GeV}^{-2}$. 
This value is in the weak-coupling region, so the diquark is unstable in the 
phase of broken chiral symmetry. Nevertheless, we shall 
show that a quark and an unstable diquark 
can form a stable baryon in this phase. 
For Figs.\ \ref{fig:re-im-pole-l}--\ref{fig:baryonstate}, we choose $G_B=10.04$ GeV$^{-1}$. 
The baryon coupling constant $G_{B}$ is actually the static approximation 
for an intermediate quark propagator in the Faddeev equation. This
approximation allows us to investigate baryon properties also at nonzero
temperature and density. 
We fix $G_B$ to obtain a baryon mass of 940 MeV in the vacuum.

In the phase diagram of Fig.\ \ref{fig:diss-diquark}, 
we choose four sets of values for temperature and
quark chemical potential, $(T,\mu_{q})=$ (0.03,0.25), (0.03,0.33),
(0.03,0.36), and (0.15,0.36), all in GeVs. 
They correspond to points A, B, C, and D. 
The red solid line separates the regions (indicated by $\chi$SB/$\chi$SR)
where chiral symmetry is broken/restored; CSC denotes the
color-superconducting phase. The blue dashed lines show the diquark
dissociation boundaries for three values of the diquark coupling
constant, $G_{D}=3.11,3.8,4.025$ (in units of GeV$^{-2}$). Below a
diquark dissociation line, the equation
$\mathrm{Re}D^{-1}(\omega,\mathbf{p}=\mathbf{0})=0$ has a real
solution $\omega$, the so-called diquark pole. The corresponding
regions in Fig.\ \ref{fig:diss-diquark} are filled with light blue,
green, and magenta color, respectively. These poles also exist in
the CSC phases, however, for the sake of clarity we choose not to
color the respective regions. The CSC phases 
are bounded by the red solid line from the left and
by the dash-dotted lines from above 
(from bottom to top for $G_{D}=3.11,3.8,4.025$, 
respectively). Note that the diquark coupling constants we have chosen here 
are in the weak-coupling or BCS regime. 
As we increase $G_{D}$, Bose-Einstein condensation of
diquarks could take place in the region below the dissociation
lines, provided the bare quark mass is nonzero 
\cite{Nishida:2005ds,Deng:2006ed,Sun:2007fc,Kitazawa:2007zs,
Brauner:2008td,Abuki:2010jq,Basler:2010xy}. 
Note that in Ref.\ \cite{Kitazawa:2007zs}, a vanishing decay width
was imposed as an additional criterion for the location of the
dissociation boundary.

The numerical results for the spectral densities are presented
in Figs.\ \ref{fig:diquark-spectral-density}--\ref{fig:diquark-spectral-density1}. 
The upper panel of Fig.\ \ref{fig:diquark-spectral-density}
shows the diquark spectral densities in the phase of broken 
chiral symmetry (point A of Fig.\ \ref{fig:diss-diquark}).
In the homogeneous limit ($\mathbf{p}=\mathbf{0}$, red solid line),
no diquark poles exist (since
$G_D=3.11\; \mathrm{GeV}^{-2}$ is too small), and the curves are smooth.
The middle panel shows the diquark spectral densities in the
phase of restored chiral symmetry, below the dissociation boundary,
but above the CSC phase (point B in  Fig.\ \ref{fig:diss-diquark}).
In the homogeneous limit, there is one sharp peak at $\omega = 0$. The
non-zero width of this peak implies that the diquark is unstable.
When temperature grows, the diquarks dissociate,
so the peak is replaced by a broad bump shown in the lower panel
(corresponding to point D in Fig.\ \ref{fig:diss-diquark}). 
In the three panels (from top to bottom) of Fig.\ \ref{fig:diquark-spectral-density1} 
we show $\rho_{R,3}$, $\rho_{I,3}$ and $\rho_{i,1/2}$ in the CSC phase
(point C in Fig.\ \ref{fig:diss-diquark}), respectively. 
For $\rho_{I,3}$ there are $\delta-$function-like peaks
in the range $|\omega|<2\Delta$, indicating stable diquarks.
For $\rho_{R,3}$ and $\rho_{i,1/2}$, these peaks attain a small width.
Also, as $\mathbf{p}$ increases, all peaks become wider. Note that the spectral
densities are not odd functions of $\omega$, because
$\mu_{q}$ is non-zero. We see that stable diquarks
only exist in the CSC region. Unstable diquark poles outside the CSC
region are actually the diquark fluctuations discussed in Ref.\ \cite{Kitazawa:2005vr}.
One can also see from the lower two panels that there are five 
Nambu-Goldstone (NG) modes 
which have poles at $\omega=0$ for zero momenta. In the lowest panel, 
there are four NG modes, i.e., the real and imaginary scalar fields with red and green color. 
In the middle panel, there is one NG mode for the imaginary scalar field 
with blue color. These existence of these NG modes is due to the validity of 
the following equations: $\frac 12 \Pi _{I/R,1/2}(0,\mathbf{0}) 
+\frac {1}{4G_D}=0$ and 
$\frac 12 \Pi _{I,3}(0,\mathbf{0}) +\frac {1}{4G_D}=0$.

In Fig.\ \ref{fig:re-im-pole-l} we show the real and imaginary parts
of the inverse retarded Greens function for baryons (positive energy component), 
again at points A,B,C, and D in the phase diagram of Fig.\ \ref{fig:diss-diquark}.
In the phase of broken chiral symmetry with $m_{q}\neq0$ and
$\Delta=0$ (point A), there are no diquark condensates or resonances
but there are stable baryon resonances: in the first panel (from top to bottom), 
we see that $\mathrm{Re}S_{B+}^{-1}(\omega_{B},\mathbf{0})=0$ has a
solution at $\omega_{B}+3\mu_{q}\approx 0.94$ GeV, i.e., close to
the rest mass of the nucleon. There is a region of
$\omega_{B}\in[-3(m_{q}+\mu_{q}),3(m_{q}-\mu_{q})]$ or
$M_{B}\in[-3m_{q},3m_{q}]$, where the imaginary part
$\mathrm{Im}S_{B+}^{-1}(\omega_{B},\mathbf{0})$ is very small
(smaller than $10^{-6}$ GeV) in the homogeneous limit. The position
is just inside this region, i.e., $M_B < 3m_q$: the baryon weighs
less than its constituents. It is therefore stable, although its constituents
by themselves are unbound, like in a Borromean state in
atomic or nuclear physics.

The second panel shows the case with diquark resonances but outside
the CSC phase (point B). There is no positive energy baryon pole in this case. 
In the region of higher temperatures and quark chemical potentials where chiral symmetry is
restored and where there are neither diquark condensates
nor resonances (point D), there are
also no baryon resonances and the absolute value of $\mathrm{Im}S_{B+}^{-1}$ is very
large. This case is shown in the third panel. In the CSC phase
(point C), there are baryon poles but with large imaginary parts, indicating
unstable baryon resonances, as shown in the fourth panel. 
This is confirmed by a broad bump in the baryon spectral density
in the fourth panel of Fig.\ \ref{fig:baryon-spec}.

The results for the baryon spectral density at different values of
$T$ and $\mu_{q}$ are presented in Fig.\ \ref{fig:baryon-spec}. 
In the first and second panels (from top to bottom), where $T=0.01,0.03$ GeV, we
observe that the baryon spectral density hardly changes with respect
to its width or peak position when varying the chemical potential
from 0.29 to 0.32 GeV. In the third panel with $T=0.05$ GeV the
peak position shows a small increase with increasing $\mu_{q}$. For
these larger temperatures, however, the width shows a dramatic
increase: the curves for $(T,\mu_{q})=(0.05,0.31),(0.05,0.32)$ GeV
are not even visible on the current scale, implying the
disappearance of the baryon resonances. For the curves still visible
at $T=0.05$ GeV, the widths are very large indicating highly
unstable baryon resonances. In the CSC phase with
$(T,\mu_{q})=(0.03,0.36)$ GeV (the fourth panel) the baryon
resonance is also quite unstable, since the peak is very low and
broad on the scale of the other panels in this figure.

%

In Fig.\ \ref{fig:baryonstate} we vary the diquark coupling constant 
in order to
investigate where the baryon is stable at $T=\mu_{q}=0$. 
We choose as $x-$axis the renormalized coupling
$G_{r}$ defined in Eq.\ (40) of Ref.\ \cite{Abuki:2006dv}. The
advantage of using $G_{r}$ instead of $G_D$ is that the existence
of stable diquark bound states is determined by the sign of $G_r$:
for $G_{r}>0$ we have diquark bound states, for $G_r<0$ they do not
exist. The shaded region in Fig.\ \ref{fig:baryonstate}
indicates where baryons are stable, i.e., where the imaginary part of the
inverse baryon propagator vanishes, or where the spectral
density may exhibit a $\delta$-function-like peak (provided the real
part also vanishes inside this region). Above
the blue curve the system is in a three-quark state (for weak
diquark coupling) or in a quark-diquark state (for strong diquark
coupling). At moderately weak negative $G_r$ the diquark is not
stable, but, as indicated by the red curve,
we obtain a stable baryonic bound state with
mass $M_{B,phys} < 3m_q$, where $M_{B,phys}$ is defined as the location
of the peak position of the baryon spectral
density. If we increase $G_r$ towards positive
values, i.e., in the range where diquarks are stable,
the pole energy of a stable baryonic bound state must lie in the range
$[-(\omega_{D}+2\mu_q)-m_q,\omega_{D}+2\mu_q+m_q]$ ($\omega_{D}$
is the energy of the diquark at $\mathbf{p}=\mathbf{0}$). The upper
boundary of this range corresponds to the blue curve which is
consequently given by $\omega_D + 2\mu_q - 2m_q$.

The threshold for stable baryons shown in Fig.\ \ref{fig:baryonstate}
by the blue curve is similar to the boundary for Efimov states
in non-relativistic cold atom physics: there, the boundary
is proportional to $-1/a_s^2$, where $a_s$ is the scattering length.
In our case, $G_r \sim a_s$, cf.\ Eq.\ (39) of Ref.\
\cite{Abuki:2006dv}. The curvature of the boundary in
Fig.\ \ref{fig:baryonstate} indeed indicates
a quadratic behavior as a function of $G_r$.

The red curve for the baryon bound state was computed with a fixed
coupling constant $G_{B}$. There are some similarities between
this state and an Efimov state. Also there, the latter cannot form, if
the two-body coupling constant is too weak, i.e., for small
negative $G_r$. On the other hand,
for a very strong two-body coupling, i.e., for small positive $G_r$,
there may be a competition between
the two-body bound state and the three-body bound state.
There are also differences to an Efimov state, for instance, in Fig.\
\ref{fig:baryonstate} the baryon bound state does not cross
the decay threshold for positive $G_r$. We perceive this
to be an artifact of a fixed
quark-diquark coupling $G_B$ in our model. In a full
calculation the quark-diquark coupling $G_{B}$ should vary
proportional to the inverse (dressed) mass of the quark exchanged
between quark and diquark \cite{Gastineau:2005wm}. Then,
$G_{B}$ will also become a function of the diquark coupling constant $G_{D}$.
A characteristics of Efimov physics is an infinite tower of
higher-lying excited states.
In order to show that they also occur in our case, we would have
to solve an eigenvalue equation for baryonic bound states.
This is a subject for future investigations.

\begin{figure}
\includegraphics[scale=0.45]{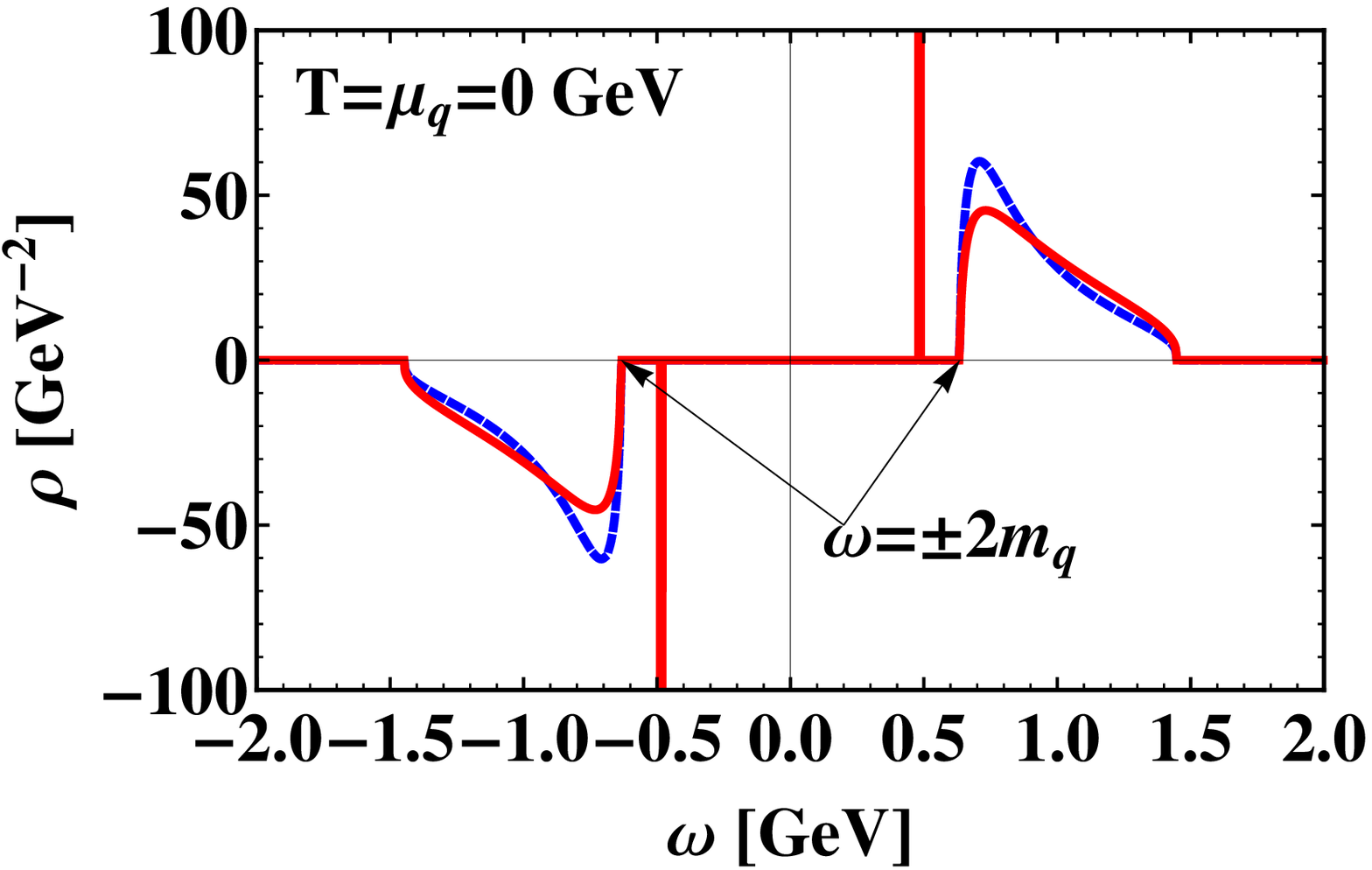}
\includegraphics[scale=0.43]{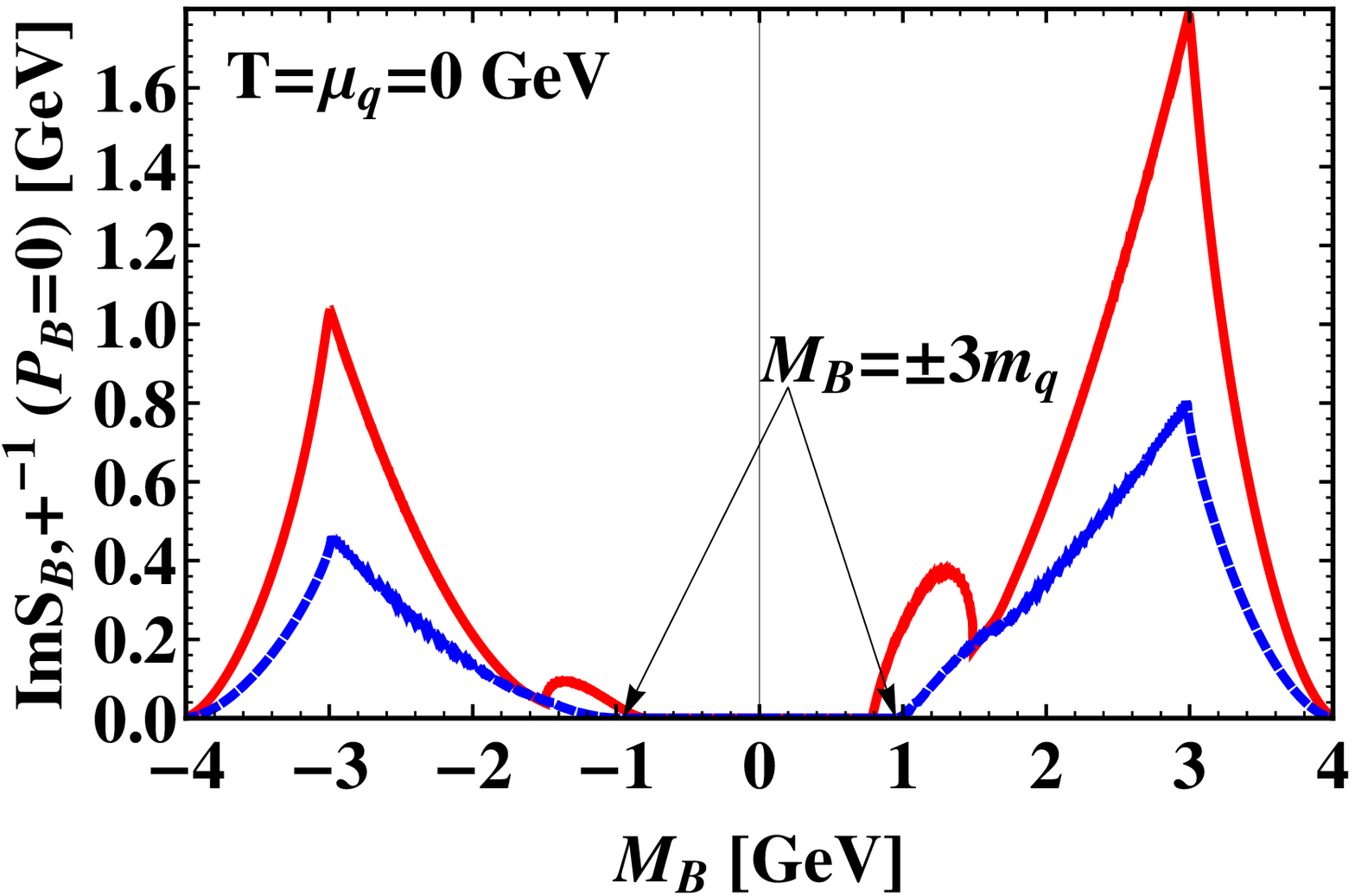}

\caption{The diquark spectral density (upper panel) and the imaginary part
of the inverse baryon propagator (lower panel). The red solid lines
are results for $G_{D}=5.95\;\mathrm{GeV}^{-2}$ and the blue dashed
lines are for $G_{D}=3.11\;\mathrm{GeV}^{-2}$. }

\label{zero-t-zero-mu}
\end{figure}

In order to see the interplay between the stable diquark and baryon
more explicitly, we present in Fig.\ \ref{zero-t-zero-mu} the diquark
spectral density and the imaginary part of the inverse baryon propagator
for $T=\mu_{q}=0$ GeV. For a strong diquark coupling (red solid line), 
one finds
two components in the spectral density, a continuous component
$\rho_{c}$ and a pole one,
\begin{equation}
\rho_{\delta}(\omega,\mathbf{p})=A(p)\delta[\omega-\omega_{p}(p)]-A(p)
\delta[\omega+\omega_{p}(p)],\label{infty-peaks}\end{equation}
where the amplitude is given by 
$A(p)=(\partial\mathrm{Re}\Pi/\partial\omega)^{-1}|_{\omega=\omega_{p}(p)}$,
and $\omega_{p}(p)$ is the energy of the pole with $p=|\mathbf{p}|$.
If the diquark coupling is weak (blue dashed line), 
only the continuous component remains,
indicating an unstable diquark. Both components are taken into
account in calculating the baryon self-energy. From the imaginary
part of the inverse baryon propagator, one finds a region
$M_{B}\in[-3m_{q},3m_{q}]$ where $\mathrm{Im}S_{B,+}^{-1}=0$ GeV
in the weak-coupling case $G_{D}=3.11\;\mathrm{GeV}^{-2}$ (blue
dashed line), where a stable baryon can be formed. In the strong-coupling 
case $G_{D}=5.95\;\mathrm{GeV}^{-2}$ (red solid line),
two additional bumps appear which overlap with 
the window $M_{B}\in[-3m_{q},3m_{q}]$.
Since nonzero
$\mathrm{Im}S_{B,+}^{-1}$ indicates unstable baryons, 
the region for stable baryons is reduced. This shows that the interplay between
pole and continuum part of the diquark spectral density is an important
ingredient in the formation of baryons. Neglecting the latter and taking
only the pole part into account misses important physics (such as
the formation of a Borromean-type stable baryon 
from an unstable diquark and a quark).

Finally, we would like to make some comparison to previous works. 
In the Faddeev approach \cite{Ishii:1995bu,Pepin:1999hs}, it is assumed
that the baryon is stable and the baryonic 
$T-$matrix has a separable form, 
which reduces the full Faddeev equation to the 
Bethe-Salpeter equation (BSE) 
for the baryonic vertex. Furthermore, for numerical simplicity 
it is also assumed that the diquark is stable.
Thus, the baryon mass can be obtained via solving an eigen-equation 
(i.e., BSE) for the baryonic vertex. 
In this approach, the effect of temperature was so far neglected
due to the increase in numerical complexity. 
An unstable diquark would also make the equation numerically hard to solve. 
Thus, so far the baryon was only treated as a stable bound state of a 
quark and a stable diquark. 
Since the baryon is stable by assumption, the properties of baryon resonances 
cannot be obtained in the Faddeev approach, 
where the baryon dissociation condition is simply realized by the condition 
that the baryon mass exceeds the sum of quark and diquark masses. 
However, this baryon dissociation condition is not correct in the CSC region 
where quarks are gapped. The correct way is to find if there 
are $\delta-$function-like peaks 
in the baryon spectral density, as done in this paper. 
Our static approximation simplifies the Faddeev equation 
to an RPA-type quasi-fermion BSE, 
so the baryon formation and dissociation at 
nonzero temperature and chemical potential 
is amenable to treatment. As we have shown, we have 
calculated the full baryonic spectral densities 
in different phases, from which the baryon 
dissociation condition is correctly obtained. 

The authors of Refs.\ \cite{Bentz:2001vc,Bentz:2002um} 
also used the static approximation in order to simplify the Faddeev equation, 
but they focus on different issues. For the diquark propagator,
they used the proper-time regularization method which 
introduces an effective confinement, but the method is not applicable to 
nonzero temperature. The diquark $T-$matrix is approximated by 
a constant term
$1/4G_{D}$ plus pole terms, which is equivalent to taking a stable diquark, 
while we employ the full spectral density of the diquark. 
There is some difference between our results 
and theirs. At low temperatures we also calculated the baryon mass 
as a function of chemical potential: we find 
only a slight decrease of the baryon mass with chemical potential, 
while they obtain a significant decrease. The reason is that we did not 
include vector mesons and thus do not obtain large baryon number densities.
Also we did not find a way of introducing confinement at nonzero temperature. 
We plan to look at these issues in a future study.
In Ref.\ \cite{Gastineau:2005wm}, the static approximation and a stable 
diquark are used. The authors considered a three-flavor NJL model, 
and the baryon mass is found 
to decrease by 25\% at normal nuclear matter density. 
We also plan to extend our model to the three-flavor case and study 
the properties of nuclear matter in the future.

In conclusion, we used an NJL-type model to compute the full diquark
propagator and its spectral density in different regions of the
phase diagram of strongly interacting matter. Baryon formation and
dissociation in dense nuclear and quark matter is then studied via
the baryon poles and spectral densities, incorporating the
previously obtained diquark propagator. We find that stable baryon
resonances with zero width are present in the phase of broken chiral
symmetry. There are no baryon poles in the chirally symmetric
phase. In the CSC phase, baryon poles exist, but they are found to
be unstable due to a sizable width. We also pointed out that
the stable baryon states found by us have some similarities
to Borromean and Efimov states in atomic or nuclear physics.

{\bf Acknowledgement.} JCW and QW thank Jian Deng for
many insightful discussions especially in the technique of
principal value integration, and
thank Lian-yi He for helpful discussions.
QW is supported in part by the '100
talents' project of Chinese Academy of Sciences (CAS) and by the
National Natural Science Foundation of China (NSFC) under Grant Nos.\
10675109 and 10735040. JCW is supported in part by China Scholarship
Council.

\end{document}